\documentclass[aps,prx,twocolumn,floatfix,superscriptaddress]{revtex4-2}
\usepackage{graphicx}%
\usepackage{color}%
\usepackage{dcolumn}% Align table columns on decimal point
\usepackage{bm}% bold math

\usepackage{amssymb,amsmath,amstext}                %%   American Physical Society math etc extensions

\usepackage{tikz,xcolor}

\usepackage{ulem}
\usepackage{latexsym}

\usepackage{soul}
\usepackage{amsmath}
\usepackage{orcidlink}

\usepackage[english]{babel}

\newcommand{\bea}{\begin{eqnarray}}
\newcommand{\eea}{\end{eqnarray}}

\def\be{\begin{equation}}
\def\ee{\end{equation}}
\def\beas{\begin{eqnarray*}}
\def\eeas{\end{eqnarray*}} 
\DeclareMathOperator{\sign}{sign}

\def\bi{\begin{itemize}}
\def\ei{\end{itemize}}

\definecolor{giergiel} {RGB}{0,118,78}

\begin{document}
\title{\textbf{Quantum-Coherent Regime of Programmable Dipolar Spin Ice} }

\author{Krzysztof Giergiel \orcidlink{0000-0003-3297-796X}}
\email{krzysztof.giergiel@gmail.com}\email{krzysztof.giergiel@csiro.au}%
\affiliation{CSIRO, Manufacturing, Private Bag 10, Clayton, VIC 3169, Australia}

\author{Piotr Sur\'owka \orcidlink{0000-0003-2204-9422}} %piotr.surowka@pwr.edu.pl
\email{piotr.surowka@pwr.edu.pl}%krzysztof.giergiel@csiro.au
\affiliation{Institute of Theoretical Physics, Wroc\l{}aw University of Science and Technology, Wybrze\.{z}e Stanis\l{}awa Wyspia\'{n}skiego 27, Wroc\l{}aw, 50-370, Poland}

\begin{abstract}
Frustrated spin-ice systems support emergent gauge fields and fractionalized quasiparticles that act as magnetic monopoles. %System
Although artificial platforms have enabled their direct visualization, access to their quantum-coherent dynamics has remained limited. %Gap/unsolved problem
Here we realize a programmable dipolar square spin-ice model using a superconducting-qubit quantum annealer, providing access to a previously unexplored quantum-coherent regime of artificial spin ice. %Claim
By implementing a direct one-to-one mapping between lattice spins and physical qubits, together with engineered extended couplings, we realize effective dipolar interactions on frustrated lattices comprising more than 400 vertices. %Implementation
Tuning transverse-field fluctuations enables us to probe the real-time dynamics of Dirac-string defects and interacting monopole plasmas. % Measurement
We observe super-diffusive monopole transport, with scaling exponents intermediate between classical diffusion and ballistic motion, indicating dynamics beyond classical stochastic relaxation and consistent with coherent propagation within an emergent gauge manifold. %Results
These results establish programmable quantum spin ice as a scalable platform for investigating fractionalized excitations and emergent gauge dynamics in engineered quantum matter. %Outlook
\end{abstract}


\maketitle

The term spin ice describes magnetic systems with an extensively degenerate ground-state manifold, characterized by a finite residual entropy analogous to that of water ice \cite{Pauling1928}. 
The minimum energy configurations obey local \emph{ice rules}, most notably the “two-in–two-out” constraint at each vertex, like the four hydrogen bonds, with two protons close and two distant from each oxygen atom. These enforce a divergence-free condition on an emergent field.
Violations of this constraint generate mobile, fractionalized excitations that carry effective magnetic charge and interact via Coulomb-like potentials \cite{Castelnovo2011}.

Artificial spin ices (ASIs) are engineered realizations of such constrained systems, originally inspired by frustrated rare-earth pyrochlores~\cite{Ramirez1999,Bramwell2001}.
These are usually two dimensional systems, including a square ASI based on the square lattice of point-like dipoles, obeying the “two-in–two-out” ice rule \cite{Lieb1967}, and a Kagome ASI, which is based on a hexagonal lattice \cite{Gartside2018}. Traditionally, artificial spin ice has been implemented using lithographically patterned nanomagnets \cite{Wang2006,tanaka2006magnetic,Farhan2019,Parakkat2021,Goryca2021,Jensen2024}, or 3D nanowire device \cite{May2021}, as well as platforms based on superconducting vortices \cite{Wang2018}, colloids \cite{Han2008,Rodriguez-Gallo2023}, and other classical realizations \cite{Miao2020MonopoleGas,Meeussen2020}.%Nisoli2013
%\cite{ortiz2019colloquium}
These systems have enabled direct imaging of monopoles and Dirac strings, but experimental access to their quantum-coherent dynamics has remained elusive. 

 %Here we study a square ASI system and see evidence of quantum fractionalized dynamics of the quasiparticle excitations that behave as effective magnetic monopoles. 
 
We build upon recent developments, which enabled the implementation of spin ice physics in reconfigurable quantum annealers, where the dipoles are represented by superconducting qubits \cite{King2021,Zhou2021}. This approach allows for not only precise control of geometry and coupling strengths, but also the ability to pin individual spins, strongly impose Gauss-law boundary conditions, and introduce topological charges at will. Unlike classical nano-magnetic realizations, the system is inherently subject to both thermal and quantum fluctuations, which can activate monopole dynamics and drive the system across its low-energy manifold, characterized by the monopole charge conservation. However, these implementations have only been able to explore classical artificial spin ices and have relied on strictly short-range, vertex-level couplings, omitting the dipolar interactions that play a central role in natural and other artificial spin ices. Dipolar interactions lift degeneracies, stabilize ordered phases, and
modify monopole kinetics through long-range energetic and entropic interaction potentials \cite{Nisoli2013}. Incorporating these interactions is therefore essential for approaching the full phenomenology of dipolar spin ice and, ultimately, regimes related to quantum spin-liquid physics.

\begin{figure*}[t]
\centering
\includegraphics[width=0.99\textwidth]{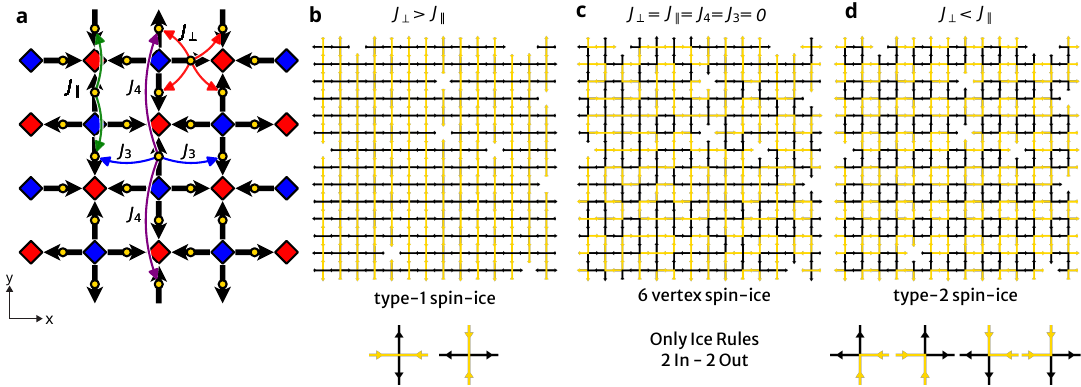}
\caption{\label{fig1} Schematic of the artificial spin ice model and its ground states.
{\bf a:} Each yellow dot represents a position of single point dipole (qubit). Straight black arrows indicate the dipole orientations. In subsequent panels dipoles that are flipped compared to this configuration are highlighted in yellow. Four types of interaction terms are included, representatives are pictured as curved arrows. %Interactions values are set to ones corresponding to truncation of two‑dimensional dipolar interactions to the first four terms.
{\bf (b-d):} 
Ice‑rule ground states obtained from annealing for different model parameters.
}
\end{figure*}

Leveraging the latest generation of superconducting-qubit quantum annealers, we realize a programmable dipolar spin-ice model, with a direct mapping between lattice spins and physical qubits and engineer effective dipolar interactions.
This allows us to study the real-time dynamics, activated by fluctuations originating from tunable transverse field, of Dirac-string defects and monopole plasmas on frustrated lattices exceeding 400 vertices. 
We access a previously unexplored quantum-coherent regime of artificial spin ice. %!!!
%The combination of programmable transverse field fluctuations and constrained spin dynamics places the system in a regime where defect motion is no longer governed by classical stochastic processes. Tracking the probability distributions of individual defects, we observe super-diffusive monopole transport, with scaling exponents intermediate between classical diffusion and ballistic motion. This behaviour indicates dynamics beyond classical relaxation and is consistent with coherent propagation of fractionalized defects within a constrained gauge manifold.
Tracking the probability distributions of motion of individual defects, we observe super-diffusive monopole transport, with THE scaling exponents intermediate between classical diffusion and ballistic motion. This behavior signals dynamics beyond classical stochastic relaxation and indicates coherent propagation of fractionalized defects within a constrained gauge manifold. We further investigate the collective evolution of many interacting defects, realizing a dynamical magnetic monopole plasma whose expansion and correlations reflect the interplay of gauge constraints and dipolar interactions. Together, these results establish programmable quantum spin ice on superconducting annealers as a new experimental platform, enabling direct studies of fractionalized excitations, entropic Coulomb interactions and emergent gauge fields in engineered quantum many-body systems. Such phenomena may ultimately find applications in electronics \cite{Heyderman2013} or low-energy computation \cite{Hon2021,Edwards2023,Caravelli2022}, with quantum collective effects making these prospects even more compelling. %%%Może jakieś lepsze wytłumaczenie tych cytowań

\section*{Methods}\label{methods}
\begin{figure*}[tb]
\centering
    \includegraphics[width=0.99\textwidth]{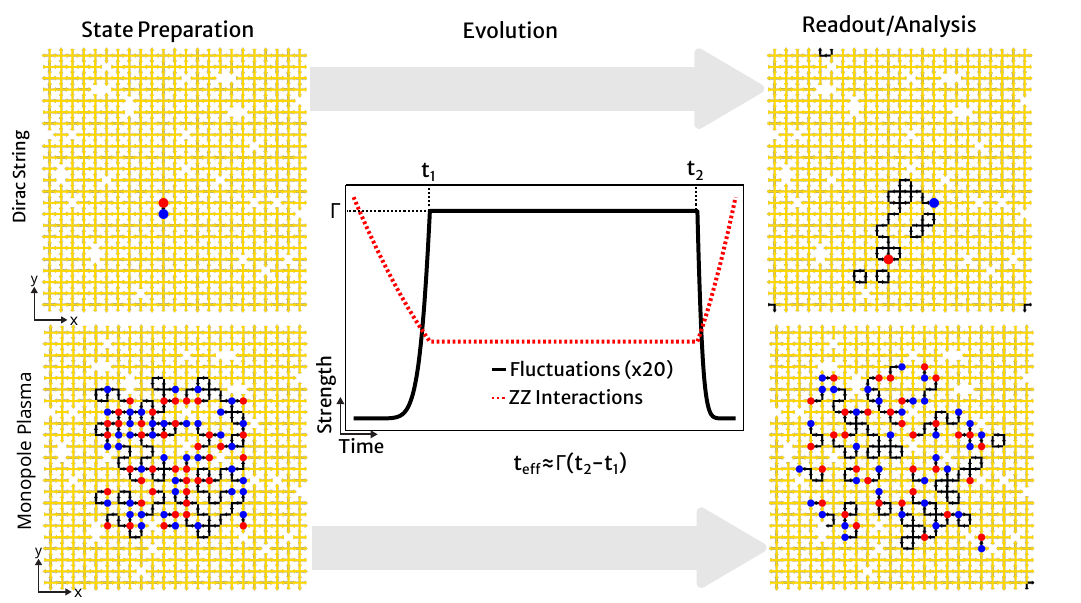}
\caption{Schematic illustration of measurement sequence. The black colors on plots here are plotted, when the dipole state is opposite to the ground state configuration (Fig. \ref{fig1}{\bf b}). Circles correspond to the vertices with $\pm2Q$ charge (3-1 in-out or 1-3 in-out). Vertices with charges $\pm4Q$ were very suppressed.  Initially the system dynamics are frozen by high values of interaction terms. We change the parameters as fast as possible decreasing the $ZZ$ interaction terms and increasing the $\Gamma X$ - fluctuation terms. The system is kept fixed $\Delta t=t_2-t_1=100~\text{ns}$ in the evolving condition. We vary the fluctuations strength $\Gamma$ to obtain different effective interactions time $t_{\text{eff}}=\Gamma\Delta t$. The $J_\perp,J_\parallel,J_3,J_4$ parameters are rescaled accordingly. The ice rules are always enforced at $45\%$ of total available value of $ZZ$ interactions. For example for $t_{\text{eff}}=100$ the energy associated with generation of pair of charges is still large $E_{2Q}=20\Gamma$, but for longer effective times ice rules are enforced less strictly.
}\label{fig2}
\end{figure*}

%Intro
We study an ASI Hamiltonian inspired by a square lattice of point-like dipoles. The
geometry, conventions, and interaction terms defining the Hamiltonian are shown in Fig.~\ref{fig1}a. Each dipole is mapped onto a single flux qubit of the D‑Wave Advantage2 quantum annealer. This mapping, described in detail in the Supplementary Material, is possible due to the increased connectivity of the newer device \cite{Advantage2}, over its predecessors.
%Annealer
The quantum annealing device consists of a network of superconducting flux qubits that implements the transverse‑field Ising Model, represented by the Hamiltonian:%\cite{Henry2014}
\begin{equation}\label{eq:ham}
\mathcal H = \mathcal J(t) \bigg( \sum_{\langle ij \rangle}J_{ij}\hat \sigma_i^z \hat\sigma_j^z + \sum_i h_i \hat \sigma_i^z\bigg) - \Gamma(t)\sum_i  \hat \sigma_i^x,
\end{equation}
where $\mathcal J(t)$ and $\Gamma(t)$ are time‑dependent schedules controlling, respectively, the strength of the classical Ising and transverse-field contributions,
$\langle ij \rangle$ indicates a sum over interacting pairs of spins $i$ and $j$.
$\hat\sigma_i^{z}$ and $\hat\sigma_i^{x}$ are the Pauli operators acting on spin $i$ in the $z$ and $x$ directions respectively. $J_{ij}$ and $h_i$ denote the programmable couplers and local bias fields, whose values remain fixed during the evolution scaled only by the global $\mathcal J(t)$ and $\Gamma(t)$. Couplings $J_{ij}$ are implemented only between physically connected qubits, corresponding to overlapping or adjacent superconducting loops. These qubits are fabricated as elongated superconducting flux loops arranged in two orthogonal layers \cite{Bunyk2014}, which enables efficient embedding of two-dimensional lattice models onto the hardware connectivity graph \cite{Boo2021}. The longitudinal($z$) Ising energy scale $\mathcal J(t)$ and the transverse‑field($x$) strength $\Gamma(t)$ are driven globally by a time dependent waveform. The local connections: $J_\perp'=J_\perp+C$ and $J_\parallel'=J_\parallel+C$ consist of a constant offset $C$, which enforces charge conservation and constrains the dynamics to the target Hilbert space, together with additional terms arising from the truncated expansion of the two‑dimensional dipolar interactions.

%Differences to previous works
Previous implementations of artificial spin-ice models on quantum annealers encoded each effective dipole using strongly coupled chains of qubits~ \cite{King2021,Zhou2021}. In this older method, a dipole flip requires a collective reversal of an entire qubit chain, constituting a higher‑order tunneling process. As a consequence, quantum fluctuations were strongly suppressed, preventing direct observation of intrinsic system dynamics. In contrast, our approach encodes each dipole in a single qubit, such that a dipole flip maps directly onto a single‑qubit transition driven by the local transverse field, enabling direct access to the system’s dynamics.
 %Before being retired faulty qubits from configuration 1.4 were apparent as they appered as localized sources of monopoles in the dynamics.
% The interactions $J_\perp, J_\parallel~\text{and}~J_4$ are achieved with available programmable couplers. 

The terms $J_\perp, J_\parallel \text{ and } J_4$ are implemented via the couplers available on the device. The interaction $J_3$ is implemented indirectly: the corresponding dipole-dipole coupling arises from the time‑averaged fast dynamics of an intermediary qubit that effectively mediates the otherwise missing connection. This approach constitutes an alternative strategy for enhancing connectivity relative to the previous qubit chain method and is particularly advantageous here, as it avoids the higher‑order tunneling that would otherwise suppress the system dynamics. Technical details of $J_3$ implementation are provided in the Supplementary Material.
%Importance of J3 J4 - should this be in Methods or main text?
The terms $J_3$ and $J_4$ are extensions of the usual fully local framework. The device permits these terms to have different signs from those typically available for $J_\perp, J_\parallel$ providing a new route to engineer more complex interacting artificial spin liquids. This capability enables access to regimes that are inaccessible in nano-magnetic systems or in previous annealing experiments. In the present work, we restrict attention to a single choice of interaction strengths corresponding to a two-dimensional dipolar interaction truncated to these four coupling types. For dipolar interaction strength $U$ the total couplings are: $J_\perp'=C+2U,~J_\parallel'=C+U,~J_3=-U,~J_4=-\frac{1}{4}U$.

%Details of implementation
We realized a lattice comprising of 23x23 vertices, with 28 vertices removed due to physical device defects that prevented implementation of the local Hamiltonian.
The measurements were run on the Advantage2 system version \emph{1.5}. Since then, device performance has since degraded, with additional qubits becoming inoperable with the current version being \emph{1.13}. We provide the complete embedding code for mapping the square‑ice model onto the Zephyr connectivity graph of the quantum annealer, together with all scripts required to reproduce the experiments, in a publicly accessible digital archive\footnote{Giergiel, Krzysztof (2026): Quantum-Coherent Regime of Programmable Dipolar Spin Ice. CSIRO. v1. Software. https://doi.org/10.25919/vraf-2p08}.  

\begin{figure*}[t]
\centering
    \includegraphics[width=0.99\textwidth]{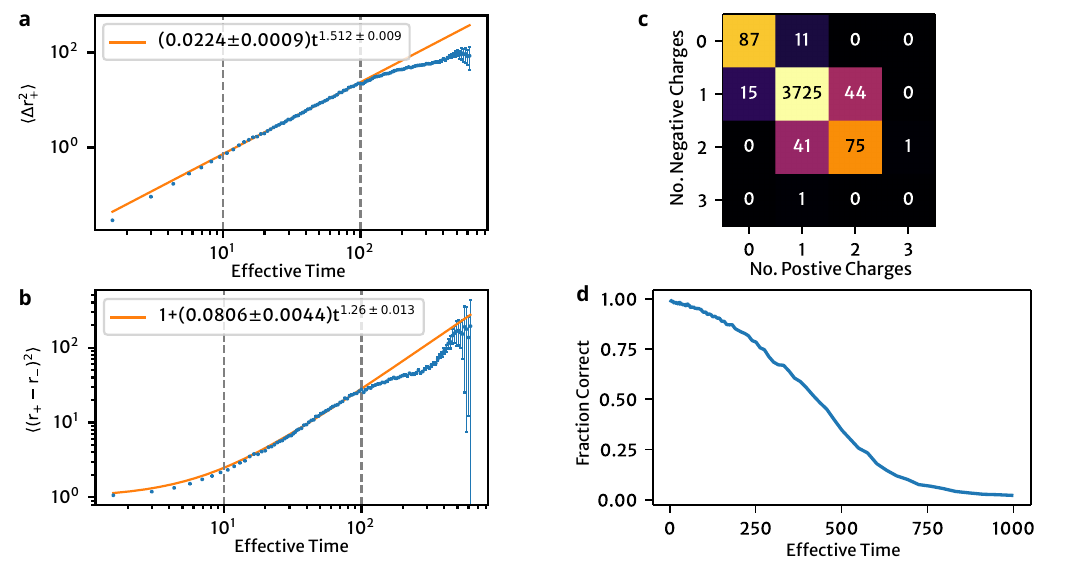}
\caption{Results of the dynamics measurement of non-interacting Dirac String defect. 
We present the log-log plots of the mean square displacement of the positive monopole at the end of the Dirac String {\bf(a)} as well as the mean square distance between ends of the string {\bf(b)}. The distance is given in the lattice units. Each point corresponds to an average of 4000 measurements over four runs of 1000 consecutive measurements, performed on different dates. The error bars correspond to the standard deviation for each point. The horizontal lines between 10 and 100 show the range of data used to fit the exponential function. The range is selected to discard the faster, transient initial dynamics and the dynamics, where fraction of samples with conserved charges is begging to fall. We recover two surprisingly different exponents about 1.5 for displacement and 1.25 for distance. {\bf(c)}: The number of observed configurations with different numbers of charges at the point $t_{eff}=100$. The 3725 corresponds to the \emph{correct} number of charges - 1 positive and 1 negative monopole - mostly corresponding to the conserved number of charges, but also including samples coming from e.g. pair annihilation and subsequent pair creation. The plot is mainly diagonal, showing the charges are predominantly annihilated and created in opposite pairs. {\bf(d)}: The fraction of samples with the correct number of charges after different durations of dynamics. The dynamics shows initially a slow non-exponential decay.}
\label{fig3}
\end{figure*}

%Measrument setup 
The previous artificial-ice platforms have demonstrated magnetic monopoles \cite{Miao2020MonopoleGas}, Dirac strings \cite{Morris2009,Mengotti2011}, and classical Coulomb phases \cite{Farhan2019}, but have lacked the programmability or dynamical control needed to tune long-range interactions or explore Z$_2$ spin-liquid behavior, particularly in the non-classical regime. We are able to prepare desired initial states, quickly increase quantum fluctuations to a desired strength and for a prescribed amount of time, and again freeze the dynamics for the readout. The states are prepared and read out in the Z basis, with the pair-wise ZZ interactions of the model Hamiltonian, and the X field supplying the fluctuations. Repeated measurements allow reconstruction of the probability distribution generated by the fluctuation driven dynamics. A schematic overview of the experimental protocol is provided in Fig.~\ref{fig2}. 

We conducted two experiments: one to observe time evolution of a single Dirac string, and a second to investigating expansion of an unconfined region of neutral magnetic monopole plasma. For the Dirac string experiments, the initial defect was positioned near the center of the lattice, maximally separated from lattice defects. In the plasma experiments, results are averaged over runs of 1000 measurements in 50 random initial configurations with a fixed initial number of monopoles seeded within the central region of the lattice.

%Calibration
To tune the parameters, we first used standard forward annealing, which generated a sample of ground states. Based on the sample we computed the vertex type statistics. The expected vertex statistics depend on the ratio of $J_\perp/ J_\parallel$ \cite{Coraux2024} and the local bias fields \cite{Goryca2021}. Representative vertex types and ice configurations obtained from annealing across different parameter regimes are shown in Fig.~\ref{fig1}. The parameters were tuned to reproduce vertex statistics consistent with pure ice‑rule behavior ($J = 0$) in the absence of bias fields, corresponding to the highly degenerate six‑vertex spin‑ice model. These parameters values correspond to the condition we describe as non-interacting.

%Experimental sequence
We chose to keep the timing of the experimental sequence fixed, using the fastest available slopes to reach the desired level of fluctuations. The system was then held at the desired level of fluctuations for $\Delta t=100~\text{ns}$. We modify only the strength of fluctuations $\Gamma$, while proportionally adjusting the ZZ terms implementing the dipolar interactions terms. This is equivalent to tuning an effective evolution time $t_{\mathrm{eff}} = \Gamma \Delta t$. The conversion to physical units is approximately $1~\text{(a.u.)} \approx 0.124$, corresponding to $0.124~\text{GHz} \times 100~\text{ns}$ at $100~\text{(a.u.)}$.    
It takes about $250~\text{ns}$ to reach fluctuation corresponding to $t_{eff}=100$. The ZZ interactions enforcing the ice rules were fixed at 45\% of total available range of the device, at the point where we observed the lowest generation and recombination rates of the monopole charges. The scaled ZZ terms implementing dipolar interactions were added on top of this baseline. In all experiments, the system dynamics were initiated from the line‑ordered state (Fig.~\ref{fig1}b), to which the desired defects were introduced (see Fig.~\ref{fig2}).
%With this calibrated interacting model and experimental protocol in hand, we next investigate how monopoles and gauge fields evolve across weakly and strongly interacting regimes.

\begin{figure*}[!htp]
\centering
    \includegraphics[width=0.99\textwidth]{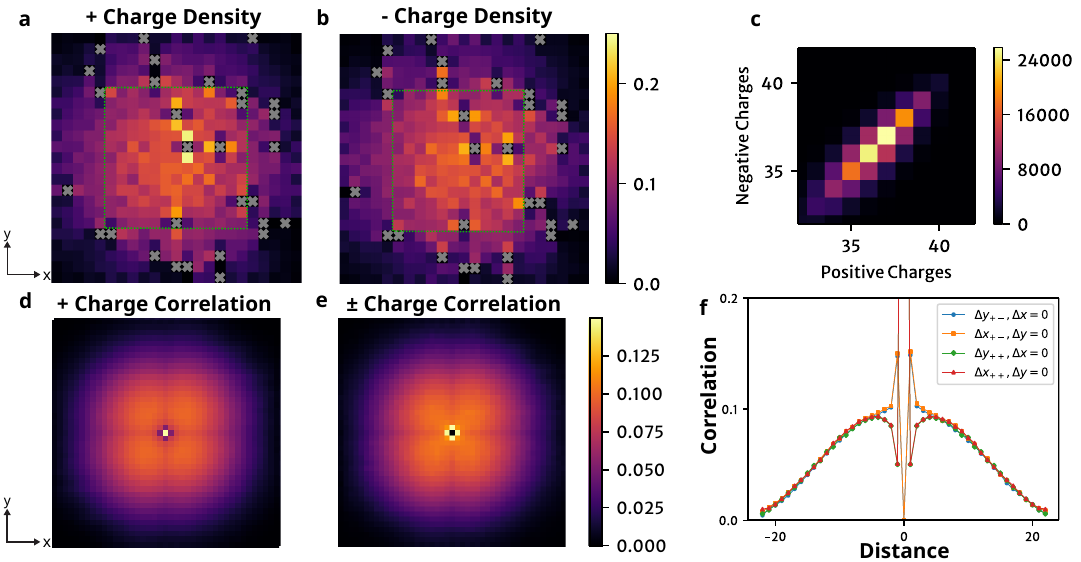}
\caption{Results of non-interacting $J=0$ monopole plasma dynamics running for $t_{eff}=100$. {\bf (a) \& (b)} The average positive and negative charge densities. The 40 initial charges of each type were placed uniformly in the areas highlighted by the green dashed rectangles. Each square pixel of the image corresponds to the charge density calculated at a single vertex. The gray $X$ symbols highlight the lattice defects - missing vertices. {\bf (c)} The distribution of experiments with given final number of positive and negative charges. {\bf (d)} The average correlation between positions of positive charges. The center of the image corresponds to the distance 0,0 and each pixel corresponds to single lattice unit distance. {\bf (e)} Similar plot for the correlations between positive and negative charges. {\bf (f)} Central vertical and horizontal cross-sections of the correlation functions shown in (d) \& (e).}
\label{fig4}
\end{figure*}

\section*{Results}\label{results}

To benchmark the dynamics we first investigate the nominally non-interacting Dirac string regime. In this configuration only the ice-rules are enforced, while dipolar interactions and residual bias field calibrated to zero within experimental resolution. Fig.~3c-d quantify charge conservation during evolution. Monopoles are observed to be predominantly created and annihilated in opposite-charge pairs, as expected from gauge constraints. At $t_{eff}=100$, approximately $93\%$ of realizations contain a single positive and a single negative monopole, corresponding to a single Dirac string. The decay of the fraction of charge‑conserving configurations is initially slow and clearly non‑exponential, indicating nontrivial dynamical constraints on defect creation and annihilation.

To characterize monopole transport, we restrict analysis to realizations that retain a single Dirac string in the interval $10\leq t_{eff}\leq 100$. Fig.~3a and 3b show evolution of mean squared displacement~(MSD) of positive monopole from its initial position, and the squared separation between the two oppositely charged monopoles. Power law fit of the form $\langle r^2 \rangle \sim t^\alpha$ yield anomalous scaling exponent $\alpha_{MSD}=1.51$ for single‑particle motion and $\alpha_{rel}=1.26$ for the relative separation.

Both exponents are smaller than the ballistic value of $\alpha=2$ expected for coherent free propagation, yet exceed than the classical diffusive value of $\alpha=1$. Moreover, the significant difference between single-particle displacement and relative separation indicates correlated motion between the two monopoles. This behavior emerges in a regime where explicit dipolar couplings are set to zero and only the ice-rule constraint is enforced. 

%TODO - Classical exponent is expected to be smaller  than 1 \cit{chakraborty2025fractionaldiffusion} 
%FROM chakraborty Moessner:\\
%a shown in the main text which gives us an exponent of 0.86.\\
%~\\
%From NISOLI:\cite{Nisoli2021EPL}\\
%\emph{In pyrochlore ice, Dusad et al. found \cite{Dusad2019IceColourExperiment} $\beta$ 1.2–1.5, suggesting that monopoles subdiffuse. In degenerate square ice, preliminary unpublished results also suggest subdiffusion.}\\
%\emph{ For degenerate square ice, a simple Glauber spin dynamics with nearest-neighbor coupling only, returns beta 1.9  close to our purely geometric estimate}\\

In a purely classical dynamics respecting the same constraint, monopole motion would correspond to thermally activated random walks along allowed spin-flip paths. Such dynamics are expected to be initially sub-diffusive, with an exponent $\alpha \approx 0.86$ based on numerical simulations~\cite{chakraborty2025fractionaldiffusion}, close to $\alpha \approx 0.85$ predicted by simple Glauber spin dynamics~\cite{Nisoli2021EPL}, before crossing over to ordinary diffusion at longer times. For systems with long-range dipolar interactions, even smaller effective exponents in the range $\alpha \sim 0.6$--$0.75$ have been reported~\cite{Dusad2019IceColourExperiment}. 

The observed super-diffusive but sub-ballistic transport in our system therefore indicates dynamics beyond classical stochastic relaxation.

The key distinguishing feature of the present platform is that fluctuations are generated by a transverse-field term 
$\Gamma X$, which induces coherent spin flips during the evolution window. The effective time parameter 
$t_{eff}=\Gamma\Delta t$ directly controls the amplitude of these quantum fluctuations. The non-diffusive scaling thus reflects monopole motion driven by controlled quantum fluctuations within a constrained Hilbert space, rather than by classical thermal hopping. In this sense, the anomalous exponents provide dynamical evidence that the defects propagate in a quantum-constrained manifold, where interference between multiple spin-flip pathways and gauge constraints modify transport relative to both classical diffusion and free ballistic motion. Our observations are consistent with a tunable, deconfined $Z_2$ gauge regime characteristic of quantum spin liquids.

We next examine collective behavior in a monopole plasma prepared with equal numbers of positive and negative charges. Fig.~4 shows results from experiments performed at $t_{eff}=100$ in the absence of explicit dipolar couplings. Initially, 40 monopoles of each sign are placed uniformly within the regions indicated in Fig.~4a–b. During evolution, charges spread across the lattice while maintaining approximate charge neutrality. The final charge-number distribution (Fig.~4c) remains strongly concentrated along the diagonal, confirming that monopoles continue to be generated and annihilated predominantly in opposite-charge pairs.

Spatial charge-density maps (Fig.~4a–b) reveal significant redistribution beyond the initial regions. The corresponding charge–charge correlation functions are shown in Fig.~4d–f. Positive–positive correlations display a short-range anti-correlation hole, while positive–negative correlations exhibit enhanced short-range attraction.

Additional illustrations in the online only Appendix Fig. A1 and A2 extend this analysis to system with different charge densities and varying interaction conditions. The initial state is similar to popping a balloon filled with plasma in a vacuum. Increasing monopole plasma density or decreasing dipolar interactions causes the cloud to expand faster. The additional effect of dipolar interactions is evident in increased short-range correlations between positive and negative charges. 

Qualitative comparison with expectations from $\mathbb{Z}_2$ lattice gauge theories suggests that the observed screening behavior, correlation decay, and absence of extended confining strings are consistent with a deconfined regime. We emphasize, however, that our measurements do not directly probe standard diagnostics such as Wilson loops or topological sector structure, and therefore do not constitute a complete identification of a $\mathbb{Z}_2$ spin liquid.

The observed dynamics provide direct access to the microscopic motion of spin-ice defects in a quantum-coherent regime of an artificial platform. While we do not yet have a quantitative theory for the measured transport exponents, the anomalous dynamics, already present in the nominally non-interacting regime, can be attributed to constraints of the Hilbert space, which generate effective entropic interactions between monopoles (see Supplementary Material for a coarse-grained derivation).

Finally, the pulsed‑evolution protocol employed here (Fig.~2) effectively probes gauge dynamics directly, offering a real‑space method to characterize emergent fields, screening lengths, and correlation functions without complications from static disorder. Although current experiments are limited in evolution time by device coherence, future improvements will enable access to longer‑time hydrodynamic regimes, where quantities such as collective modes and damping rates of a quantum monopole plasma may be directly measured.

{\textbf{Supplementary materials}}\\
%3.9 MB pdf file: SM.pdf
%Supplementary Material: Programmable Dipolar Quantum Spin Ice on a Superconducting Qubit Annealer\\
1.2 MB Dirac\_String\_Density\_Evolution.mp4 file:\\
Animation showing the evolution of the positive charge density used to generate Fig.~3a. Plots include only the data with correct charge counts and are normalized by the number of shots (4000).\\
1.2 MB Dirac\_String\_Correlation\_Evolution.mp4 file:\\
Animation showing the evolution of the charge correlation density used to generate Fig.~3b. Plots include only the data with correct charge counts and are normalized by the number of shots (4000).\\
7.5 MB Configurations.mp4 file:\\
Movie of some of the configurations measured in the calibration of the energy scale of charge conservation enforcement. Initial state is the Dirac string shown in Fig. 2, configurations are measured after dynamics were run for $t_{eff}=100$.
\begin{acknowledgments}
We thank Sherry Mayo for a careful reading of the manuscript and valuable comments. We thank Nilotpal Chakraborty for bringing the references about sub-diffusive scaling to our attention. P.S. acknowledges the funding from the Polish National Science Centre (NCN) Sonata Bis grant 2019/34/E/ST3/00405. Access to the D-Wave system was funded by the CSIRO Quantum Technologies Future Science Platform.
\end{acknowledgments}

\section*{Author Contributions}
K.G. conceived the study, implemented the models, performed the experiments, and carried out the data analysis. K.G. also prepared the figures and videos.
P.S. contributed to the conceptual development of the models and experiment design, provided physical interpretation of the results, and advised on alternative theoretical approaches. 
Both authors discussed the results, refined the analysis, and contributed to the final version of the manuscript and supplementary material.

\section*{Data availability}
The data used to generate the animations and Fig.~3 are included in the archive along with the code. Requests for raw, shot‑by‑shot data should be directed to K.G. (krzysztof.giergiel@gmail.com).
The codes for this study are available at \href{https://doi.org/10.25919/vraf-2p08}{doi.org/10.25919/vraf-2p08}. The archive contains all code used to run the experiments via the D-Wave Ocean platform, as well as the scripts for data analysis and plotting, along with the necessary data.

\begin{figure*}[t]
\centering
    \includegraphics[width=0.99\textwidth]{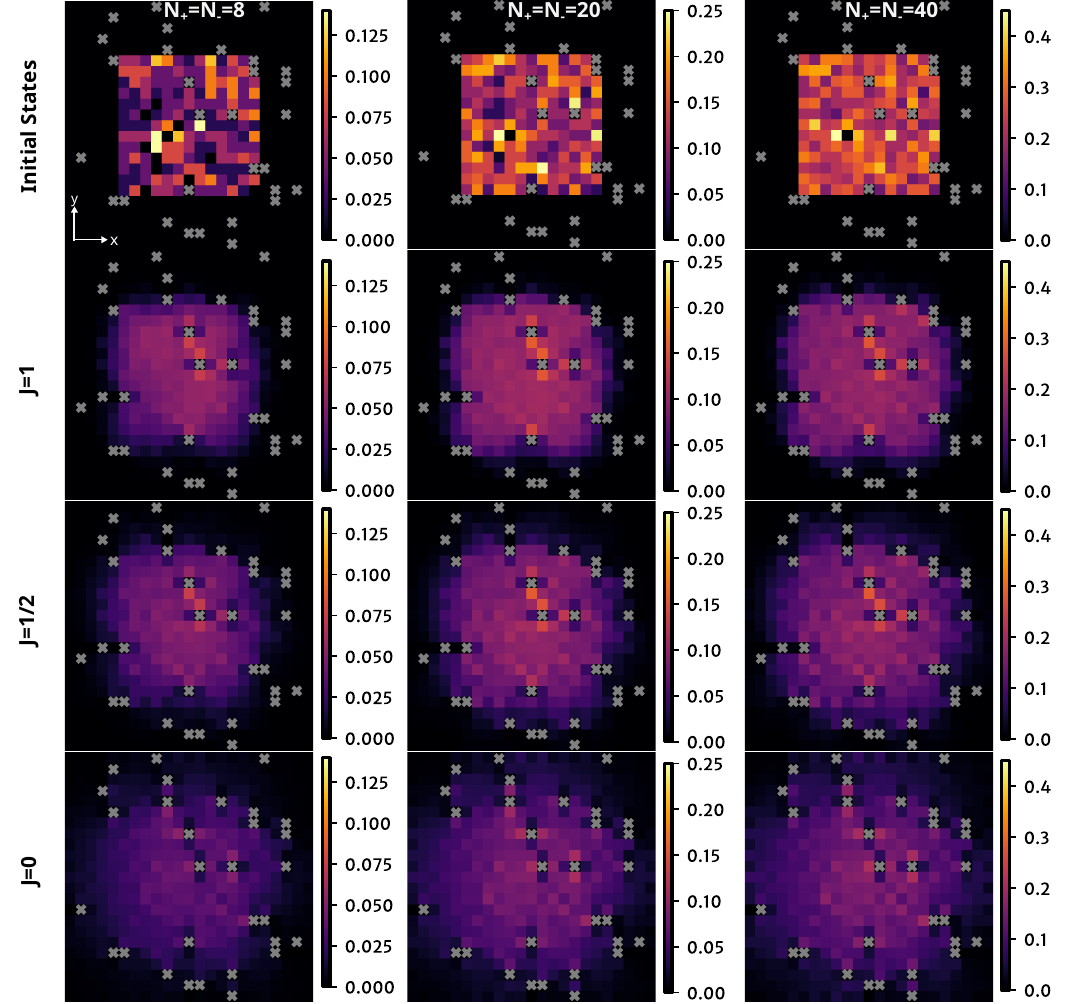}
\caption{Array of positive charge density plots. Top row depicts the average of the 50 initial states. Subsequent rows show average of states after $t_{eff}=100$ of evolution, at different 2D dipolar interaction strengths. Different columns correspond to different number of initial charges.
}
\label{fig5}
\end{figure*}

\begin{figure*}[t]
\centering
    \includegraphics[width=0.99\textwidth]{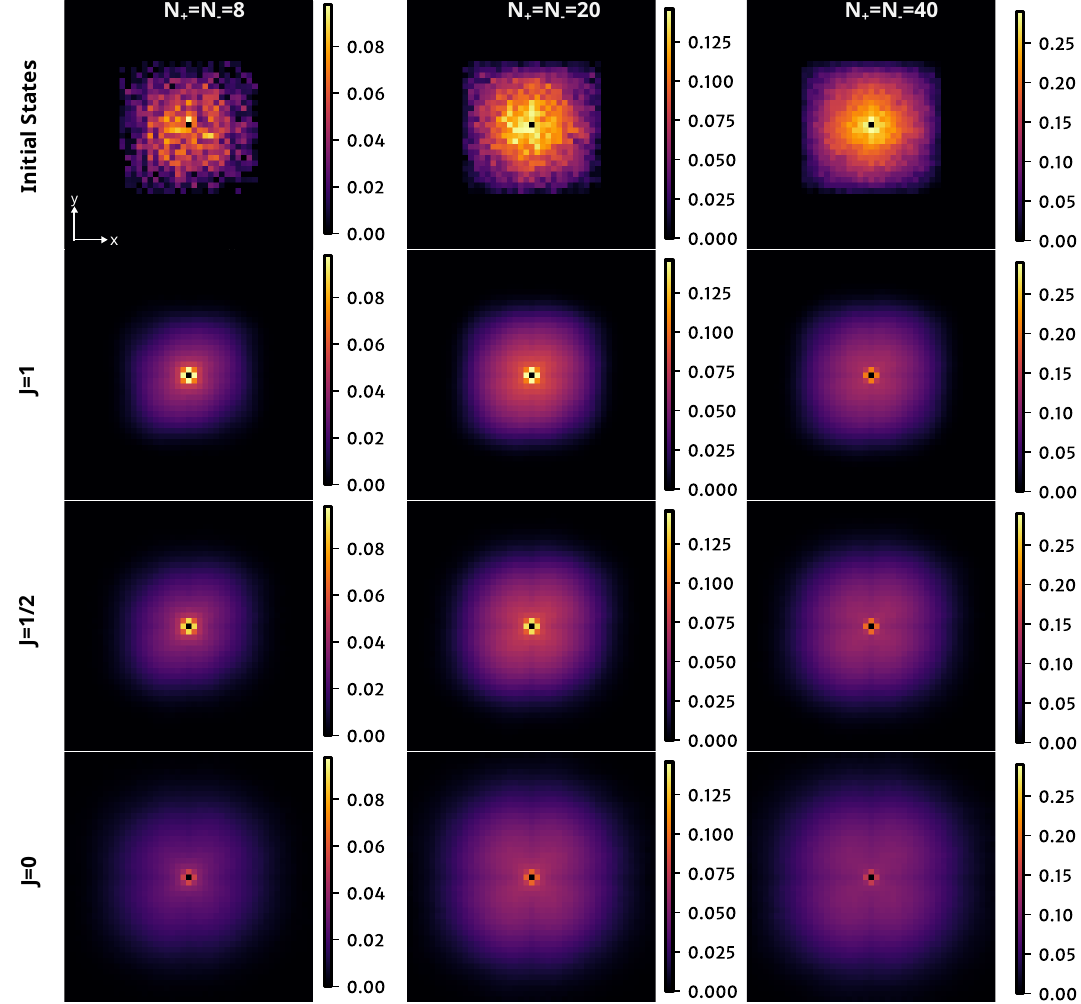}
\caption{Array of correlation plots of positive and negative charges positions. Top row depicts the average of the 50 initial states. Subsequent rows show average of states after $t_{eff}=100$ of evolution, at different 2D dipolar interaction strengths. Different columns correspond to different number of initial charges.}
\label{fig6}
\end{figure*}

\appendix

\section{Emergent Gauge Field and Magnetic Charges in Artificial Spin Ice}

\subsection{Spin configuration and coarse-grained field}

In artificial spin ice systems spins reside on the links of a lattice and obey local ice-rule constraints. Each link $i$ carries an Ising variable
\begin{equation}
\sigma_i = \pm 1 ,
\end{equation}
aligned along the bond direction.

We define $\sigma_i = +1$ when the spin points from odd(red) sublattice site to site even (blue) sublattice, and $\sigma_i = -1$ for the opposite orientation, see main text Fig.~1.

A coarse-grained emergent magnetic field can be constructed from the microscopic spins as
\begin{equation}
\mathbf{B}_{\mathrm{eff}}(\mathbf r)
=
\sum_i
\sigma_i\,\hat{\mathbf e}_i
\delta(\mathbf r-\mathbf r_i),
\end{equation}
where $\hat{\mathbf e}_i$ is the unit vector along link $i$ and $\mathbf r_i$ denotes the center of the link.

\subsection{Ice rule as a divergence constraint}

The ice rule imposes that the net spin flux entering each vertex $v$ vanishes,
\begin{equation}
Q_v = \sum_{i\in v} \sigma_i = 0 .
\end{equation}

In the continuum limit this condition becomes
\begin{equation}
\nabla\cdot\mathbf B_{\mathrm{eff}}(\mathbf r)=0 ,
\end{equation}
which is analogous to Gauss’s law for magnetism. Thus the ice-rule manifold corresponds to configurations of a divergence-free emergent field.

\subsection{Magnetic charges from ice-rule violations}

When the ice rule is violated the divergence becomes non-zero. A vertex with net spin flux
\begin{equation}
Q_v = \sum_{i\in v} \sigma_i
\end{equation}
acts as a localized magnetic charge. In the continuum description this produces
\begin{equation}
\nabla\cdot\mathbf B_{\mathrm{eff}}(\mathbf r)
=
\sum_v Q_v\,\delta(\mathbf r-\mathbf r_v).
\end{equation}

These excitations correspond to emergent magnetic monopoles.

For example, on the square lattice:

\begin{itemize}
\item $2$-in/$2$-out: $Q_v = 0$ (neutral vertex)
\item $3$-in/$1$-out: $Q_v = +2$ (positive monopole)
\item $1$-in/$3$-out: $Q_v = -2$ (negative monopole)
\item $4$-in or $4$-out: $Q_v = \pm 4$ (higher-charge excitation)
\end{itemize}

Monopoles can propagate through the lattice via successive spin flips, generating effective magnetic currents.

\section{Entropic interaction between monopoles}

Even in the absence of dipolar interactions, monopoles interact through an
entropic Coulomb interaction arising from the constraint
\begin{equation}
\nabla\cdot\mathbf B = \rho_m ,
\end{equation}
where $\rho_m$ is the magnetic charge density.

The coarse-grained probability distribution for the field is
\begin{equation}\label{eq:gaussian_weight}
P[\mathbf B] \propto  e^{-F/T} \propto
\exp\!\left[
-\frac{K}{2 v_{\mathrm{cell}}}
\int d^d r\, |\mathbf B(\mathbf r)|^2
\right],
\end{equation}
where $F$ is the corresponding coarse-grained free-energy functional, and 
$T$ is the temperature. The coefficient 
$K$ is a dimensionless stiffness (entropic stiffness in the Coulomb-phase description) $v_{\mathrm{cell}}$ is the volume of a lattice cell \cite{castelnovo_debye-huckel_2011}.
Let us introduce a positive monopole charge $+Q$ at position $\mathbf r_1$ and a negative charge $-Q$ at position $\mathbf r_2$. 
The divergence constraint becomes
\begin{equation}
\nabla \cdot \mathbf B =
Q\,[\delta(\mathbf r-\mathbf r_1)-\delta(\mathbf r-\mathbf r_2)] .
\end{equation}

Minimizing the functional \eqref{eq:gaussian_weight} under this constraint
gives the saddle-point field $\mathbf B_\star = -\nabla\phi$ with

\begin{equation}
\nabla^2\phi = -\rho_m .
\end{equation}
Substituting the saddle-point field yields the entropic free energy
\begin{equation}
\frac{F_{\mathrm{ent}}}{T}
=
\frac{K}{2 v_{\mathrm{cell}}}
\int d^d r\, |\mathbf B_\star(\mathbf r)|^2 .
\end{equation}

For two charges $Q_1$ and $Q_2$ separated by $r$, the interaction part becomes
\begin{equation}
\frac{W_{\mathrm{ent}}(r)}{T}
=
\frac{K}{v_{\mathrm{cell}}} Q_1 Q_2 G(r),
\end{equation}
where $G(r)$ is the Green's function of the Laplacian. In two dimensions
\begin{equation}
G(r)
=
-\frac{1}{2\pi}
\ln\!\left(\frac{r}{a}\right),
\end{equation}
where $a$ is a microscopic cutoff. Finally,
\begin{equation}
W_{\mathrm{ent}}(r)
=
-\,k_B T\,
\frac{K}{2\pi v_{\mathrm{cell}}}
Q_1 Q_2
\ln\!\left(\frac{r}{a}\right).
\end{equation}
Qualitative comparison with expectations from $\mathbb{Z}_2$ lattice gauge theories suggests that the observed screening behavior, correlation decay, and absence of extended confining strings are consistent with a deconfined regime. We emphasize, however, that our measurements do not directly probe standard diagnostics such as Wilson loops or topological sector structure, and therefore do not constitute a complete identification of a $\mathbb{Z}_2$ spin liquid.

The observed dynamics provide direct access to the microscopic motion of spin-ice defects in a quantum-coherent regime of an artificial platform. While we do not yet have a quantitative theory for the measured transport exponents, the anomalous dynamics---already present in the nominally non-interacting regime---can be attributed to constraints of the Hilbert space, which generate effective entropic interactions between monopoles (see Supplementary Material for a coarse-grained derivation).

\begin{figure*}[t]
\centering
% \begin{overpic}[width=0.48\textwidth]{ConnectionsNNNV3.png}
%        \put(0,72){\textbf{a}} 
%  \end{overpic}
\includegraphics[width=0.48\textwidth]{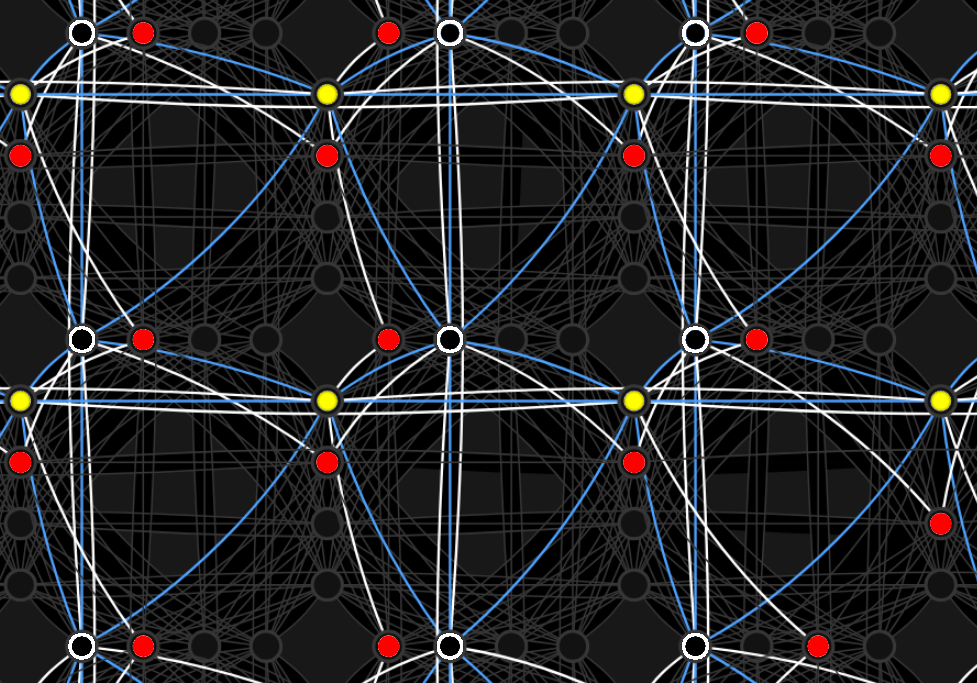}
\includegraphics[width=0.48\textwidth]{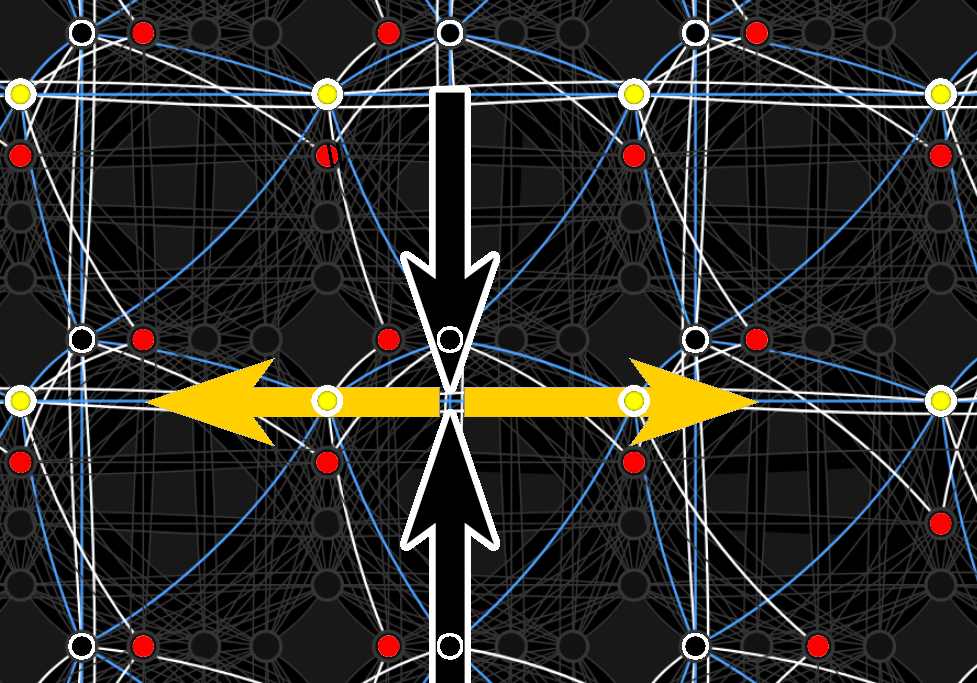}\\
\includegraphics[width=0.48\textwidth]{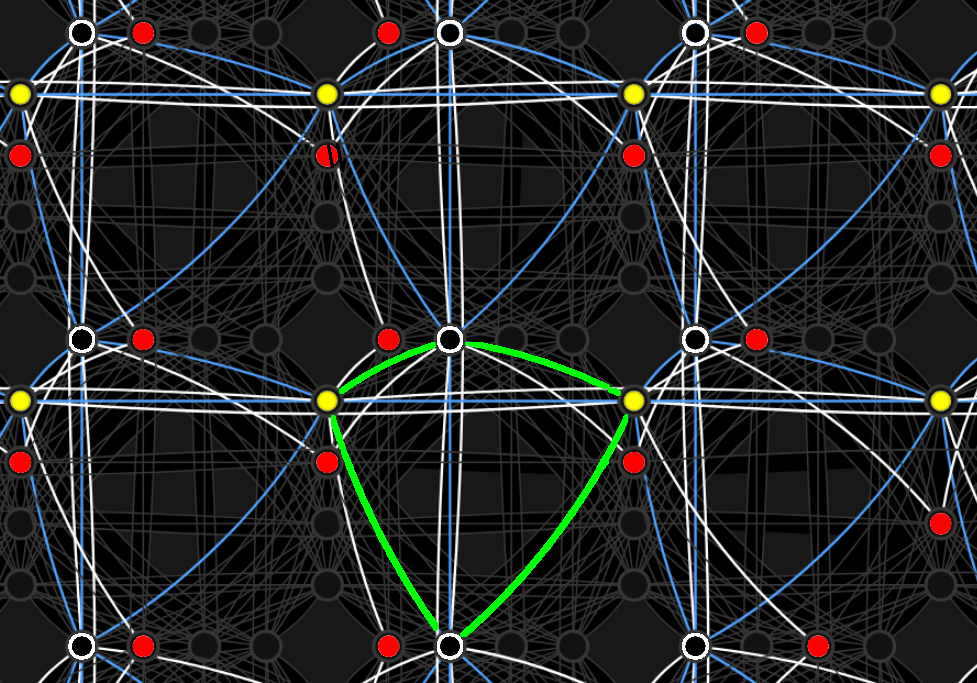}
\includegraphics[width=0.48\textwidth]{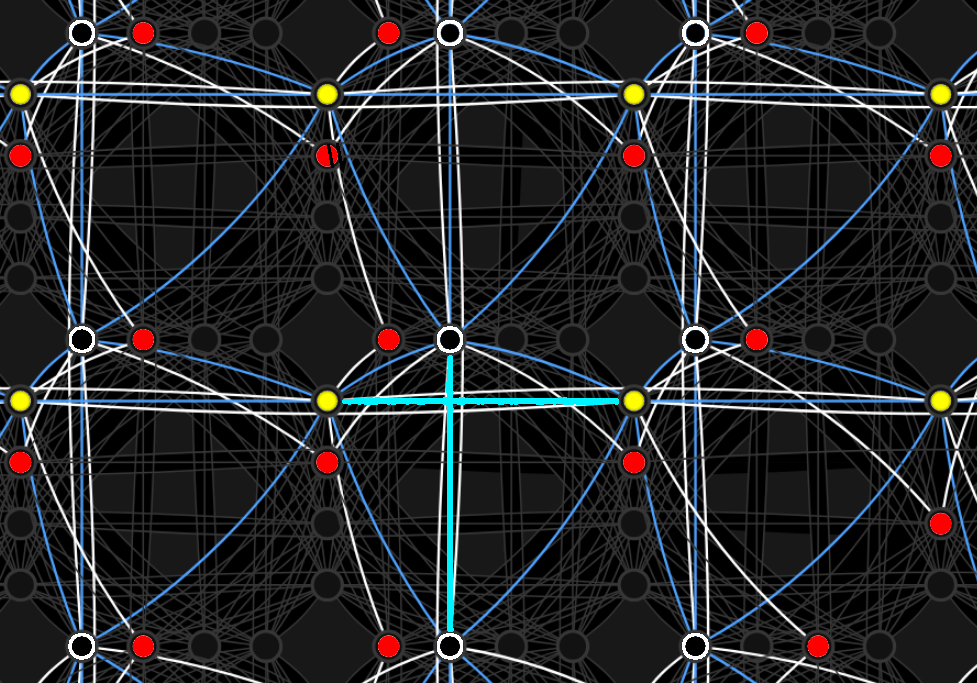}\\
\includegraphics[width=0.48\textwidth]{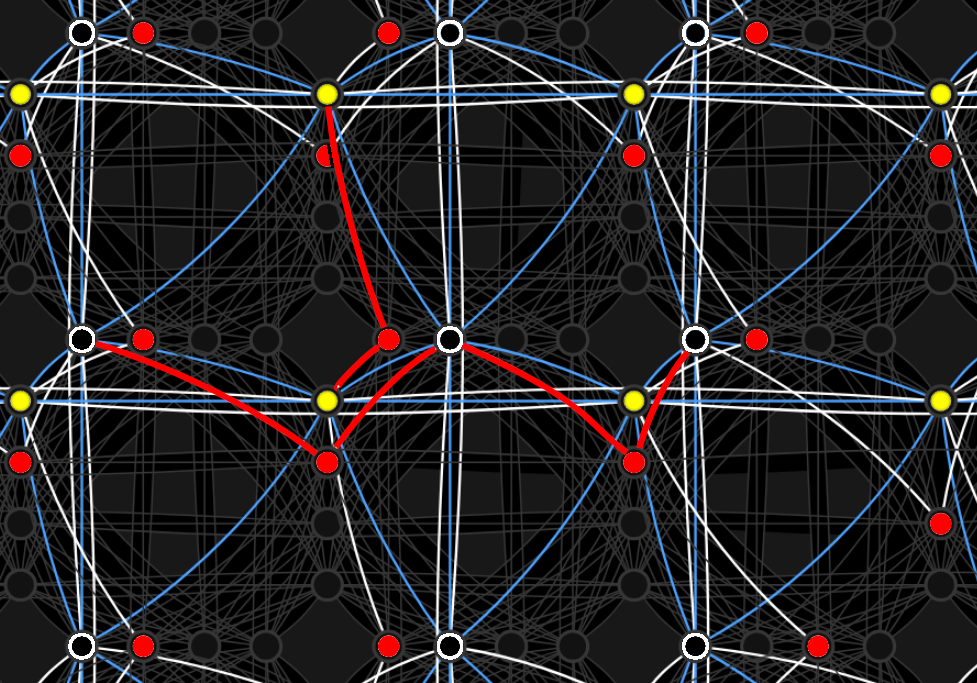}
\includegraphics[width=0.48\textwidth]{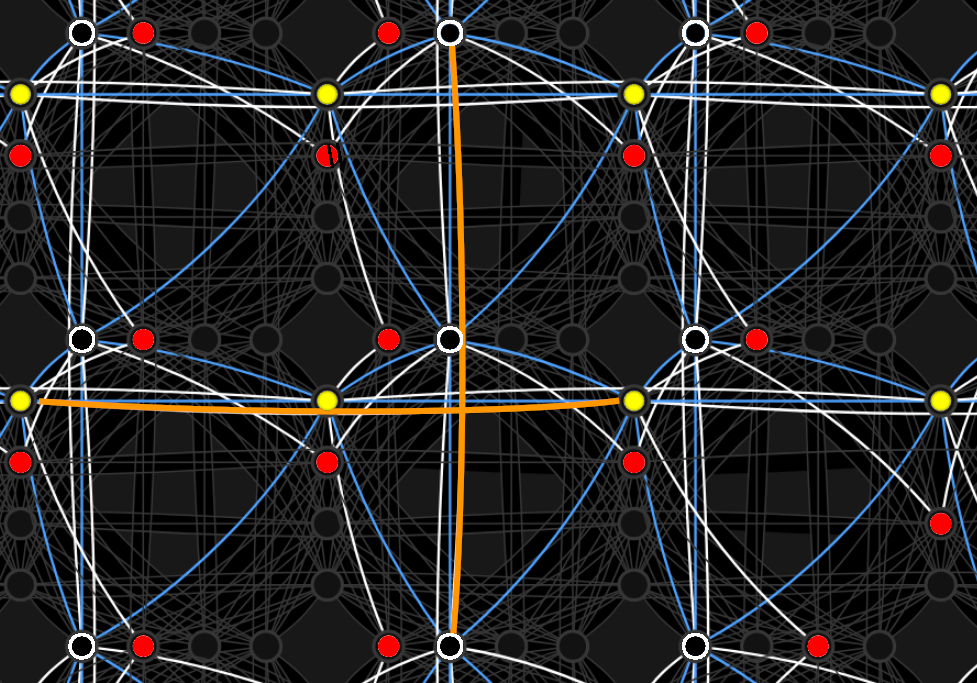}\\
\caption{{\bf Top Left:}  Schematic view of the layout positions - dots, and connections - curved lines, of the flux qubits on the DWAVE device. Coloured in dots are the qubits used to represent the ASI, rest is grayed out to indicate they are unused. Yellow and black correspond to the horizontal/vertical dipoles in the spin ice state, as in Fig. 1 in the main text. The red dots, represent the qubits mediating the $J_3$ interactions.
Blue and white color of connections represents the connections sign.  
{\bf Top Right:} A unit cell associated with a single vertex superimposed on the structure. {\bf Center Left:} Highlighted in green: representative $J_\perp$ connections. {\bf Center Right:} Cyan - $J_\parallel$ connections. {\bf Bottom Left:} Two horizontal and one vertical $J_3$ connections highlighted in red.  {\bf Bottom Right:} One horizontal and one vertical $J_4$ connections highlighted.}
\label{figS_J3}
\end{figure*}

\section{Dipolar Interaction Between Spins}

In artificial spin ice the spins also interact through dipolar forces. For dipoles $\mathbf p_1$ and $\mathbf p_2$ separated by $\mathbf r$, the interaction energy in two dimensions can be written as
\begin{equation}
U_{2D}(\mathbf r)
=
\frac{\mathbf p_1\cdot\mathbf p_2}{|\mathbf r|^2}
-
2\,
\frac{(\mathbf p_1\cdot\mathbf r)
(\mathbf p_2\cdot\mathbf r)}
{|\mathbf r|^4}.
\end{equation}

These interactions modify the microscopic energetics but do not change the entropic origin of the monopole interaction derived above.

\begin{figure*}[t]
\centering
\includegraphics[width=0.99\textwidth]{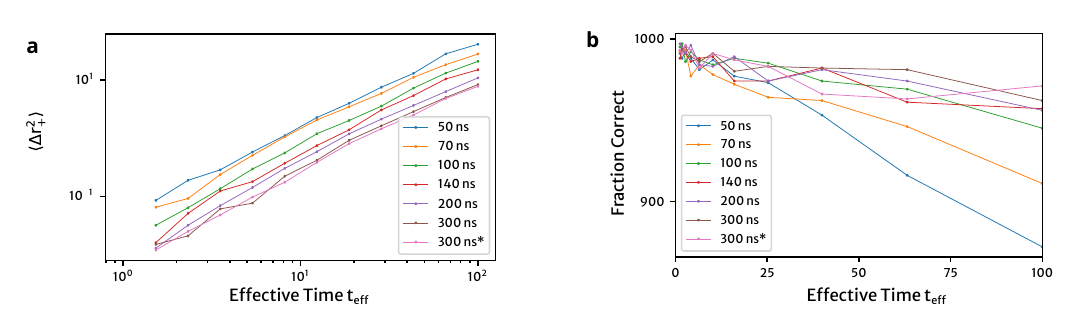}
\caption{Comparison of results of the dynamics measurement of non-interacting Dirac String defect for different dwell times $\Delta t$ -see legend. The $*$ denotes a run, where correction to $\Gamma$ from the expected effects of parameters ramps are included. {\bf(a)} We present the log-log plots of the mean square displacement of the positive monopole at the end of the Dirac String. The distance is given in the lattice units. Each point corresponds to an average of 1000 measurements from a single run. {\bf(b)}: The fraction of samples with the correct number of charges after different durations of dynamics. The dynamics shows initially a slow non-exponential decay. }
\label{FigGamma}
\end{figure*}
\section{Implementation of long range interaction term \texorpdfstring{ $J_3$}{J3} }\label{SuppJ3J4}
We can use additional connecting qubit-spin between two spins which are not natively connected.
This qubit will be driven with much higher X field ensuring its dynamics are fast.
The dynamics of our $Z_2$ model are slow - hence the connecting qubit is experiencing slow - annealing dynamics. It's contribution to energy follows the ground state configuration (for this qubit).
The fast Hamiltonian looks as follows:
\bea
H_c=J_1 \sigma_1 \sigma_c + J_2 \sigma_2 \sigma_c + H \hat{X}\sigma_c,
\eea
where $\hat{X}$ is Pauli-x. Let $|J_1|=|J_2|=J$. There is energy difference, which prefers ferro or anti-ferro configurations of $\sigma_1$ and $\sigma_2$ decided by the sign $|J_1\cdot J_2|$. 
The energy difference:
\bea
\Delta E = \sqrt{H^2+J^2}-|H|\approx \frac{J^2}{2H}\text{~ for small J}
\eea
induces the connection in the slow Hamiltonian:
\bea
H_{slow}=4 \sign(J_1J_2)\Delta E \sigma_1\sigma_2.
\eea
For this to be true the condition the dynamics of $c$ have to be much faster then dynamics of model spins $\Gamma \hat{X} \sigma_1$, $H>>\Gamma$. Practically this is achieved on the DWAVE machine by setting the annealing offsets of the control qubits to large negative values. 

\section{Testing the \texorpdfstring{$t_{eff}=\Gamma \Delta t$}{teff=Gamma Dt}.}
This test is meant to check that the dynamics depend primarily on the $t_{eff}$, and not the thermal fluctuations. The effect of thermal fluctuations is expected to scale with $\Delta t$, while coherent explanation would depend only on $t_eff$. In Fig.~\ref{FigGamma} we repeat the Dirac String experiment with limited statistics (1000 samples per point, single measurement run).  We additionally check that inclusion of a correction for the expected dynamics during the initial as final ramps by substraction of the time integral of the fluctuations during these ramps: $\int \text{d}t \Gamma(t)$ and adjusting the constant $\Gamma$ is not producing a correction visible in the measurement results. 

Decreasing $\Delta t$ requires and increase of $\Gamma$ to keep the same $t_{eff}$, and this seems to come with a cost of leaving the conserved charge sub-space as evidenced in Fig.~\ref{FigGamma}b, while also facilitating quicker charge movement shown in Fig.~\ref{FigGamma}a. Crucially results show that the mean squared displacement scaling in $t_{eff}$ is the same across different dwell times $\Delta t$. This results provide some evidence that the main driving force behind the spin ice defect's movement is coherent quantum evolution.

\begin{figure*}[t]
\centering
\includegraphics[width=0.99\textwidth]{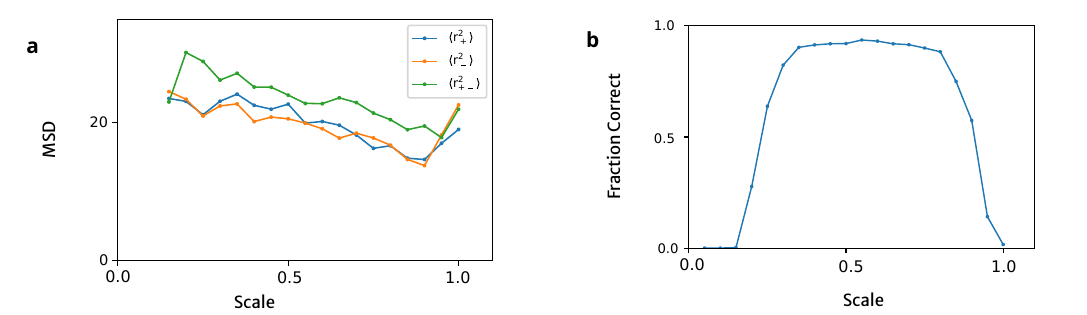}
\caption{Parameter scan over the strength of enforcing the Gauss' Law  $J_\perp'=J_\parallel'=C$. {\bf a}: The Mean Square Displacement of the positive and negative charges and the Mean Square Distance between the positive and negative charges (labeled $+-$). The plot includes only shots, where the number of charges was identical to the initial single Dirac String state.  {\bf b}: The fraction of shots, where the number of charges was identical to the initial single Dirac String state.}
\label{FigScale}
\end{figure*}

\section{Impact of the energy scale of charge conservation enforcement}

We tested to check that the dynamics do not depend significantly on the choice of energy scale, which enforces the Gauss condition around the chosen value. We scanned over the strength of the local connections: $J_\perp'=J_\parallel'=C$ enforcing the charge conservation condition, see Fig.~\ref{FigScale}. We observe the transport is slightly inhibited by increasing the energy scale $C$ and when set to high we likely couple outside the qubit Hilbert space of the lowest energy states, to higher flux states. We selected the value 0.45 as the second lowest value point in our parameter scan, where the plotted fraction of charge conserved runs became stable. The animation attached to the article shows a recent sample of the configurations , when scaling the strength of the Gauss-Law terms. These were obtained from Advantage2 version 1.11 (corresponding to later date than the results shown in main text). The breaking of charge conservation is evident there, as at low values additional defects are being generated, and at the high values increased annihilation of the monopole pair is observed, as relaxation to the defect free spin ice is occurring.

\section{Calibration}
During the calibration we discovered a preference of dipole alignment in the y-direction. The out-of the box correlation matrix between different spin directions coming into the vertex was as follows:
\[
\begin{pmatrix}
dir&up & left & right & down\\
up&1 & -0.25 & -0.37 & -0.37 \\
left&-0.25 & 1 & -0.37 & -0.37 \\
right&-0.37 & -0.37 & 1 & -0.25 \\
down&-0.37 & -0.37 & -0.25 & 1
\end{pmatrix}
\]
Out of the box, the statistic revealed asymmetry, between top and bottom layer of superconducting loops, which are laid out in orthogonal directions (up/down and left/right), which correspond to the directions of the dipoles. We tuned the local bias fields and the bias of $J_\perp$ and $J_\parallel$, which resulted in the balanced correlation matrix:
\[
\begin{pmatrix}
1 & -0.33 & -0.33 & -0.33 \\
-0.33 & 1 & -0.33 & -0.33 \\
-0.33 & -0.33 & 1 & -0.33 \\
-0.33 & -0.33 & -0.33 & 1
\end{pmatrix}.
\]
At that point, after calibration, the average vertex statistics obtained from standard forward annealing look as follows:
\[
\begin{vmatrix}
Vertex  & a_{hor} & b_1 & b_2 & b_3 & b_4 & a_{ver}\\
Counts & 88915 & 82466 & 82911 & 82716 & 82362 & 76546 \\
Expected & 92938 & 78281 & 78281 & 78281 & 78281 & 92938 
\end{vmatrix}
\]
\[
\setcounter{MaxMatrixCols}{11}
\begin{vmatrix}
-4Q & -2Q & -2Q &-2Q&-2Q & 2Q & 2Q & 2Q & 2Q &4Q\\
0   & 429 & 395 & 354 & 384 & 399 & 399 & 359 & 365 & 0\\
0   & 0   & 0   & 0   & 0   & 0   & 0   & 0   & 0   & 0
\end{vmatrix}
\]
where the expected row corresponds to the counts based on the numerical results for $20x20$ lattice with open boundary conditions presented in \cite{Coraux2024}, scaled to the number of (non-defective) vertices in our lattice (499). The numbers describe the number of vertices out of 499000 (based on the 1000 states of the 499 active vertices). The $a$, vertices correspond to the two directions of the type-1 vertices, and $b$ are four different type-2 vertices. The type-3 and -4 vertices are named according to their charges ($|2Q|$,$|4Q|$). We run small Dirac string experiments with and without this bias calibration and observed the same scaling of the charge movement exponents, within the shot noise. This suggests the main results should be robust even without the calibration step.

Obtained vertex statistics show that we are close to an extensively degenerate six-vertex model square ASI. This is in contrast to $J_\perp \neq J_\parallel$ case of the generalized F-model. Anisotropic experiments are usually the simpler to realize, as for example the natural dipolar interactions in nano-magnetic platforms prefer the line order ~\cite{Wang2006}, but even bias in the opposite direction results in sub-extensive entropy~\cite{Perrin2016}. The highly degenerate regime $J_\perp = J_\parallel$ is more interesting as it maximizes the number of ground states giving rise to the entropic effects that are characteristic of ices.

\bibliography{bibliography}% bib file

\end{document}